\documentclass[conference]{IEEEtran}
\IEEEoverridecommandlockouts
\usepackage{cite}
\usepackage{amsmath,amssymb,amsfonts}
\usepackage{algorithmic}
\usepackage{graphicx}
\usepackage{textcomp}
\usepackage{xcolor}
\usepackage{listings}
\usepackage{comment}
\usepackage{subcaption}
\usepackage[hyphens]{url}

\newcommand{\revision}[1]{#1}

\def\BibTeX{{\rm B\kern-.05em{\sc i\kern-.025em b}\kern-.08em
    T\kern-.1667em\lower.7ex\hbox{E}\kern-.125emX}}
\begin{document}

\title{Towards Automatic Model Completion: from Requirements to SysML State Machines}

\author{\IEEEauthorblockN{Maria Stella de Biase}
\IEEEauthorblockA{\textit{Dip. di Matematica e Fisica} \\
\textit{Universit\`a della Campania  ``L. Vanvitelli''}\\
Caserta, Italy \\
mariastella.debiase@unicampania.it}
\and
\IEEEauthorblockN{Stefano Marrone}
\IEEEauthorblockA{\textit{Dip. di Matematica e Fisica} \\
\textit{Universit\`a della Campania  ``L. Vanvitelli''}\\
Caserta, Italy \\
stefano.marrone@unicampania.it}
\and
\IEEEauthorblockN{Angelo Palladino}
\IEEEauthorblockA{\textit{Aerospace Business Unit} \\
\textit{Kineton srl}\\
Napoli, Italy\\
angelo.palladino@kineton.it}
}

\maketitle

\begin{abstract}
Even if model-driven techniques have been enabled the centrality of the models in automated development processes, the majority of the industrial settings does not embrace such a paradigm due to the procedural complexity of managing model life cycle. This paper proposes a semi-automatic approach for the completion of high-level models of critical systems. The proposal suggests a specification guidelines that starts from a partial SysML (Systems Modeling Language) model of a system and on a set of requirements, expressed in the well-known Behaviour-Driven Design paradigm. On the base of such requirements, the approach enables the automatic generation of SysML state machines fragments. Once completed, the approach also enables the modeller to check the results improving the quality of the model and avoiding errors both coming from the mis-interpretation of the tool and from the modeller himself/herself. An example taken from the railway domain shows the approach.
\end{abstract}


\section{Introduction}\label{sec:Introduction}

The increasing complexity of critical systems requires a greater and greater level of software dependability while market pressure pushes down products time-to-market. To this aim, it is crucial to shorten software development periods without sacrificing quality. The best way to reduce fatal errors is to take preventive measures. Since the definition of requirements is the first fundamental step in the development process, according to the software engineering V-Model~\cite{b3} (mainly used in safety-critical systems development~\cite{b13}), it is necessary to sift through the clarity, correctness and consistency of the requirements in order to eliminate certain types of errors in advance.

In this research, we follow the well-known Model-Based Development (MBD)~\cite{b12}, which explored different aspects of the software development and of the above-mentioned V-model. This notwithstanding, the MBD approaches mainly focused in the past years on the ``lower-level" activities (e.g., code generation, testing). On the other hand, industries do not prefer big changes in their assessed processes due to the necessity of managing multi-years ongoing projects; instead, they prefer small improvements that easily adapt to existing practices.

To this aim, this paper describes the first idea of an ongoing research whose primary aim is mixing the contribution of both humans and automatic tools in a requirement engineering process. The goal is to build a tool that, starting from a partial SysML model of a system and from a set of textual requirements, is able to complete the model by creating some SysML model fragments to integrate in the global one. We are aware that this goal is very ambitious and then there are some working hypotheses for the research:
\begin{itemize}
    \item The requirements are defined according the well-known Behavioural Driven Development (BDD)~\cite{b2} principle and using the \texttt{Given-When-Then} paradigm;
    \item The requirements are also structured according to guidelines considering different ways (requirement templates) of writing a BDD requirement;
    \item SysML model is itself built according to a specific guideline where components and their main states are partially defined. 
\end{itemize}

Under these preliminary conditions, the first goal is to build an automatable process able to: (1) read requirements and choose the most suitable model fragments for them; (2) generate state machines' transitions capturing the behaviour described by the requirement; (3) annotate the requirement with a set of SysML annotations to improve traceability and to ease the modeller in verifying the final result of the model.

The fulfillment of this concrete objective will be possible by enhancing traditional writing processes with formal grammars and using Artificial Intelligence (AI) and Natural Processing Language (NLP) algorithms. SysML profiling methods will also be followed, too. 

The industrial impact of this research is high since one of the most critical obstacles in the adoption of a full model-driven process in industrial settings is constituted by the physiological disalignment between models and other ``runnable" artefacts. In this way, the model would be at the center of the system development process since its early phases; automation would enable the model as a ``living object" of the process, and the traceability information enables explainability and interpretability that are crucial for critical systems, \revision{mitigating the traditional disinclination of international standards as ISO 26262, DO-178B, and DO-178C to AI and NLP adoption}. One of the ever-changing critical systems is the railway domain.

Railway is one of the safest modes of transport\footnote{\url{https://international-railway-safety-council.com/safety-statistics/}} and the safety is achieved by properly managing its infrastructure.
To this aim, the European Train Control System (ETCS) is constantly improving performance and safety of the standard: currently, the ETCS Level 3 (ETCS-L3) \cite{b15} is under development and different research projects are still assessing the preliminary set of the standard requirements. A concrete example taken from the railway signalling systems will show the presented approach.

This paper is structured as follows: Section~\ref{sec:Relatedworks} presents an overview of the state of the art, issues and solutions related to requirements definition and domain modelling. Section~\ref{sec:Methodology} describe the steps to be followed, supported by a small example. Section~\ref{sec:techdiscuss} discusses technical concerns as well as reporting the status of the research.

\section{Related Works}\label{sec:Relatedworks}
Requirement definition is a crucial and delicate stage. The requirement writing process generates a detailed description of the system, indeed requirements express \revision{what and how much well system functions should perform \cite{b3}}.
According to the expressed features of the system, it is possible to categorize the requirements (e.g., Functional, Safety, etc.). 
Requirements are usually defined in Natural Language (NL), although intuitive may generate omissions or ambiguities~\cite{b6}. Over the years, many works developed solutions and techniques to improve the definition of system requirements. 
All the revised works consider the attributes of acceptability as starting point (see Table~\ref{tab:tab_attribute}). They propose different approaches aimed at improving the requirement definition, e.g., grammar-based~\cite{b7}, pattern-based~\cite{b1,b6}, formal models-based~\cite{b9,b10,b11}.
\begin{table}[htbp]
    \centering
    \begin{tabular}{p{3cm}|p{4.5cm}}
    \hline
    \textbf{Attribute} & \textbf{Description} \\
    \hline
    Clear and Consistent & Readily understandable \\
    Consistency & Terms used are consistent \\ 
    Correct & Does not contain error of fact \\
    Feasible & Can be satisfied by technologies, etc. \\
    Singular & One actor-verb-object requirement \\
    Verify & Provable at the level of the architecture \\
    Without ambiguity & Only one interpretation makes sense \\ [1ex]
    \hline
    \multicolumn{2}{c}{\textbf{For pairs and sets of Requirement Statements}} \\ \hline
    Absence of conflicts & Not in conflict with other requirements \\ 
    Absence of redundancy & Each requirement specified only once \\ [1ex] \hline
    \end{tabular}
    \caption{Attributes of acceptability~\cite{b3}}
    \label{tab:tab_attribute}
\end{table}

The survey in~\cite{b8} defines two categories of requirements formulation approaches: \textit{direct} and \textit{statistical machine translation}. The first category is based on the usage of a context free grammar to translate NL text in a temporal model. The second one deals with the writing requirement process, considering them as a result of a language translation subject to probabilistic rules. Under the \textit{direct approaches} umbrella, which uses patterns to collect requirements, it is possible to find the BDD, a set of methods developed by Dan North in the agile development framework~\cite{b2}. Cucumber~\cite{b1} is the software that supports BDD technique providing Gherkin\footnote{\url{https://cucumber.io/docs/gherkin/}}. The latter is a language that allows the definition of system requirements in NL by means of a patterns. 
The authors in~\cite{b10} analysed different \textit{temporal logic formulae extraction} techniques and then derived the common steps. The \textit{extraction of syntactic information} is one of them, and it is applied in order to highlight entities in the texts with appropriate tags. This stage of NLP is called Named Entity Recognition (NER) and it is used to label each word of the text with its grammatical category (e.g., noun, verb, etc.). This step is carried out in~\cite{b11} for sentence classification purpose. 

\revision{The integration of modelling and AI is a field of growing interest, as it is witnessed by some emerging workshops on the topic\footnote{https://mde-intelligence.github.io/} and by a recent special issue on Software and System Modeling \cite{b19}. Other relevant works include: \cite{b14}, where an NLP-based assistant is proposed to facilitate the domain modelling providing completion suggestion for the partial model under construction; \cite{b16}, where an interactive bot proposes to the modeller configurations for the model under construction. Other works develop NLP-based techniques to extract domain models from text \cite{b17,b18}. 
}


\section{Methodology}\label{sec:Methodology}
\begin{figure*}[htbp]
    \centerline{\includegraphics[width=0.8\textwidth]{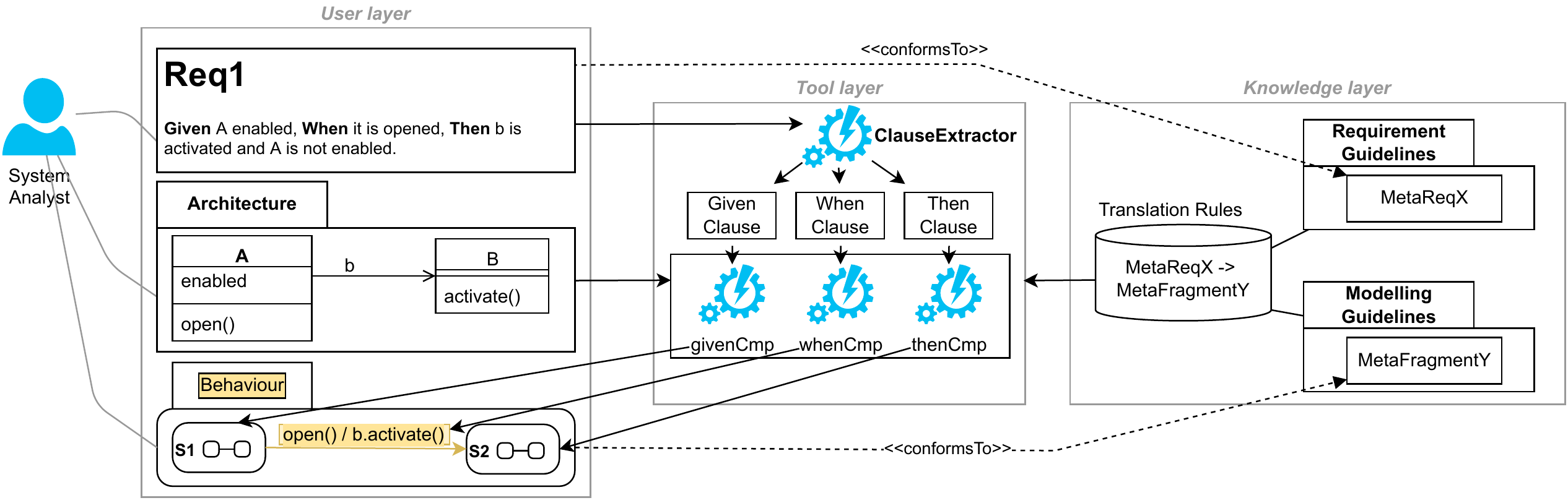}}
    \caption{\revision{A Conceptual Architecture}}
    \label{fig:workflow}
\end{figure*}

The goal of this research work is to provide a semi-automatic support to the software development process focusing on a first fundamental step: the Requirement Engineering phase. Looking at Fig.~\ref{fig:workflow} describes the proposed approach at-a-glance \revision{by representing a conceptual architecture of the components to develop}. Three layers are presented to separate and organize the elements of the approach.

\paragraph*{User layer} it considers the parts of the approach that are in charge of the \textit{System Analyst}. The underlying hypothesis is that the requirement engineer has to provide both a set of requirements expressed according to the BDD paradigm and a partial system model (i.e., \textit{Architecture} in the figure). The discussion on the size of such initial model is in Section~\ref{sec:techdiscuss}. Here, we make the hypothesis that such an initial model considers a complete Class Diagram (CD)/Block Diagram (BD) containing the complete classes/blocks, their relationships, methods, attributes, ports, flows, etc. The initial model also considers a partial State Machine Diagram (SMD) --- the \textit{Behaviour} in Fig.~\ref{fig:workflow} --- where only states are defined without any indication of the transitions\footnote{The part of such SMD is in white in the figure.}.

\paragraph*{Knowledge layer} the approach is based on the \textit{detect-and-translate} idea, i.e., the detection of requirement patterns inside the specification, with the consequent definition of a model fragment that best represents the patterns in SysML. 
To formalise such idea, the Knowledge layer mainly contains the \textit{Translation Rules} knowledge base mapping meta-requirements (\textit{MetaReq}) --- the requirement patterns --- and the resulting SysML fragment (\textit{MetaFragment}). From a user perspective, this knowledge base ``generates" two guidelines, one for writing requirements and one for modelling behaviours in SysML, respectively Requirement and Modelling Guidelines. The schemes of the \textit{MetaReq}s are discussed in Section~\ref{sec:techdiscuss}.

\paragraph*{Tool layer} to support the automation of the approach, a toolchain will be developed, considering the following steps. First, a \textit{ClauseExtractor} is in charge of getting the parts of a BDD requirement. To this aim, a simple parser based on a formal grammar will be developed considering the main keywords of the Gherkin language (i.e., \textbf{given}, \textbf{when}, \textbf{then}, \textbf{and}, \textbf{or}). Then the clauses are separated and a simple Abstract Syntax Tree (AST) is generated from the requirement. Three different tools --- namely \textit{givenCmp}, \textit{whenCmp}, \textit{thenCmp} --- receive as input: the AST of the requirement, the initial SysML model, the knowledge base, improving as output the SMD by creating transitions and annotating them with triggers, conditions and actions.

\subsection{A Railway Signalling example}

\begin{figure}[htbp]
    \centering
    \includegraphics[width=0.4\textwidth]{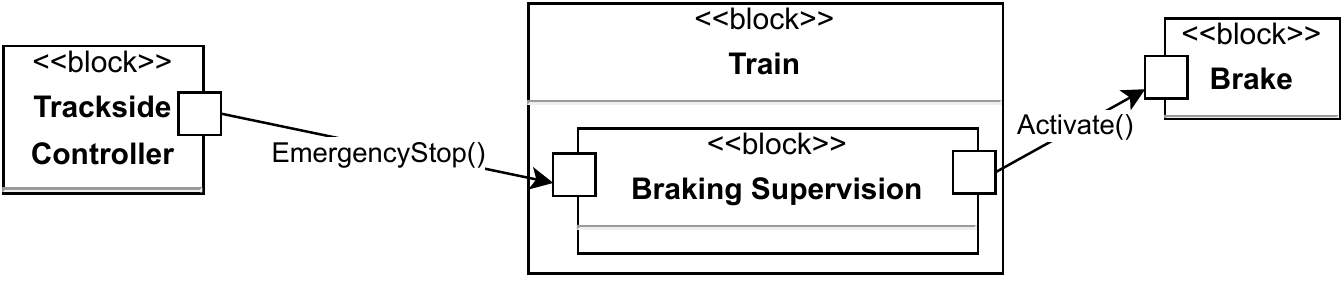}
    \caption{Train System Block Diagram}
    \label{fig:blockDiagram}
\end{figure}

\begin{figure}[htb]
     \centering
     \begin{subfigure}[b]{0.2\textwidth}
         \centering
         \includegraphics[width=\textwidth]{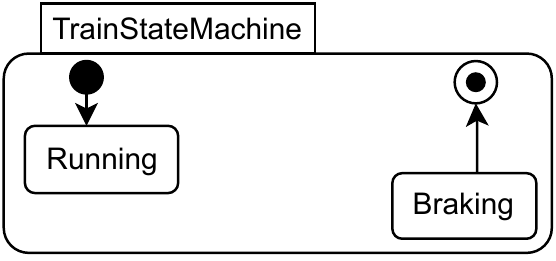}
         \caption{Incomplete SMD}
         \label{fig:incompleteSM}
     \end{subfigure}
     \hfill
     \begin{subfigure}[b]{0.2\textwidth}
         \centering
         \includegraphics[width=\textwidth]{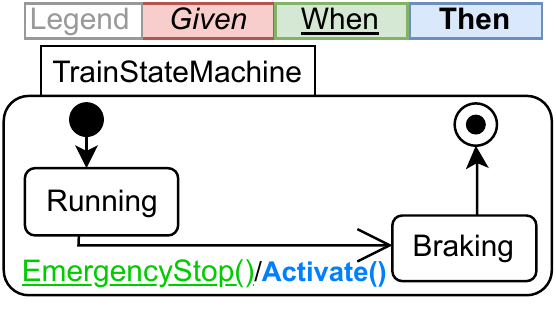}
         \caption{Complete SMD}
         \label{fig:completeSM}
     \end{subfigure}
\caption{Train SMDs}
\end{figure}

Let us consider the example of the partial system model, in the railways domain depicted in Fig.~\ref{fig:blockDiagram} to show the final aim of the automation to define. A trackside controller communicates with the train by sending the EmergencyStop() message in case of immediate braking. A related safety requirement is:
\begin{center}
\fbox{\begin{minipage}{23em}
\textbf{Given} a Train in running, \textbf{When} the Braking Supervision receives an Emergency Stop Message, \textbf{Then} the Braking Supervision activates the Emergency Brake and goes in braking.
\end{minipage}}  
\end{center}
By also considering the partial SMD of the train reported in Fig.~\ref{fig:incompleteSM} where only two states are present (i.e., Running and Braking), the proposed approach is aimed at enrich such SMD then obtaining the one of Fig.~\ref{fig:completeSM}. This is done by adding the transition from the Running to Braking states. Such a transition has as trigger, the \textit{EmergencyStop()} signal, and an action \textit{Activate()} which causes a signal to be sent to the Brake blocks.

\section{Technical Discussion}\label{sec:techdiscuss}
Some technical details on how to design and implement the approach are presented in the following.
\paragraph*{Structure of the Knowledge Base}

Even if the structures of the knowledge base, of the MetaReqs and of the MetaFragments are not defined yet, here a preliminary discussion about their formalisation is reported. Table~\ref{tab:requirementAndMap} reports on the upper part, the requirement considered in the railway example (Requirement) and MetaReq structure (MetaReq).

\begin{table}[]
    \centering
    \begin{tabular}{p{4.1cm}| p{3.5cm}}
        \hline
        \textbf{MetaReq} & \textbf{Requirement} \\
        \hline
        Given $<<$Block as context1$>>$ in $<<$State as starting$>>$, When $<<$Block as context2$>>$ receives $<<$Signal as event$>>$, Then $<<$Block as context3$>>$ $<<$Signal as operation$>>$ to the $<<$Block as context4$>>$ and goes in $<<$State as final$>>$. & Given a Train in running, When the Braking Supervision receives an Emergency Stop Message, Then the Braking Supervision activates the Emergency Brake and goes in braking.\\
        \hline
        \multicolumn{2}{c}{\textbf{Elements Matching}} \\
                \hline
        \textbf{Role} & \textbf{Value} \\
        \hline
        context1 (Block) & Train \\
        starting (State) & Running \\
        context2 (Block) & Braking Supervision \\
        event (Signal) & EmergencyStop() \\
        context3 (Block) & Braking Supervision \\
        operation (Signal) & Activate() \\
        context4 (Block) & Brake \\
        final (State) & Braking \\
        \hline
    \end{tabular}
    \caption{Pattern, text and elements matching}
    \label{tab:requirementAndMap}
\end{table}

\begin{figure}[htbp]
    \centering
    \includegraphics[width=0.3\textwidth]{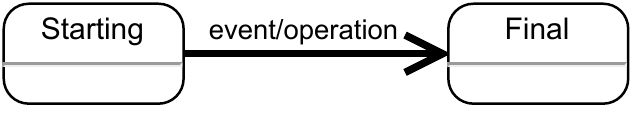}
    \caption{Clauses-SMD Mapping}
    \label{fig:metafragment}
\end{figure}

The MetaReq includes some of the keyword of the BDD requirement structure and a list of \textit{matching element patterns}, whose syntax is given by the notation \textit{$<<$metclass as role$>>$}. 
A concrete requirement fits a MetaRequirement if for each element pattern \textit{$<<$metclass as role$>>$} there exists a word that corresponds to a model element, instance of the \textit{$<$metclass$>$}, that plays the given \textit{$<role>$} in the requirement.
For example, in ``Given a Train in running", the word \textit{Train} corresponds to the element pattern $<<$Block as context1$>>$ because Train is a Block in the SysML model. So, Train plays the role of \textit{context1}. Roles are then used to fill the corresponding MetaFragment, that is depicted in Fig.~\ref{fig:metafragment}.

The lower part of the Table~\ref{tab:requirementAndMap} reports the matching of the elements on the considered railway domain example.

Another consideration is related to the matching mechanism. Two ``extreme" criteria are possible: 1) to define a rigid Backus-Naur Form (BNF) for the MetaReq, forcing the System Analyst to use a fixed schema, 2) to fully use NLP techniques to match the concrete requirement with the possible MetaReq. The flexibility/effectiveness trade-offs between these two solutions is an ongoing part of the research.

\paragraph*{On the Initial Model}
The work presented so far starts from relaxed initial conditions. The automatic approach presented in this paper starts from a System Under Test (SUT) model comprising blocks, detailed with attributes, methods and ports if necessary, furthermore states of the main blocks are also considered. The idea is to develop this research through incremental steps starting from a basic but incomplete SMD model until the construction of such a SMD model from scratch. In this way, it will be possible to assess the results obtained at each step. Future works would also consider the possibility to built automatically the rest of the SUT model.

\begin{figure}[htbp]
    \centering
    \includegraphics[width=0.45\textwidth]{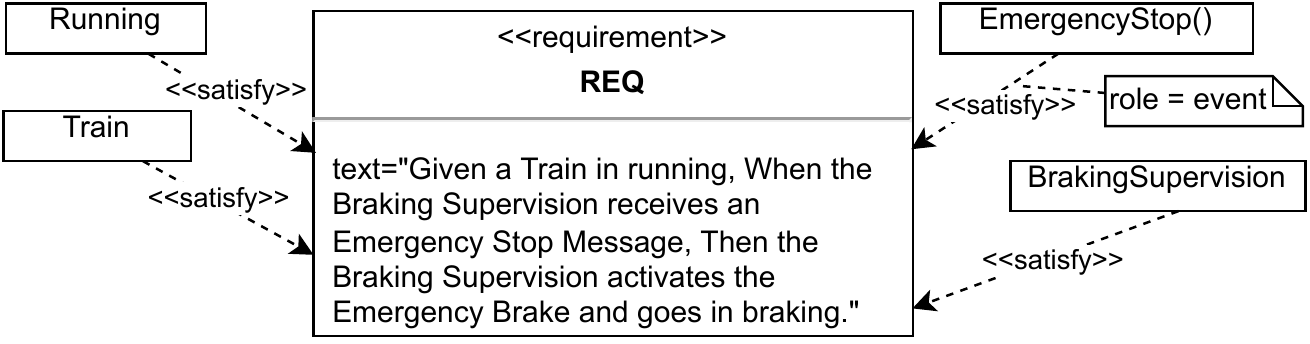}
    \caption{Requirement Traceability}
    \label{fig:traceability}
\end{figure}

\paragraph*{On the Traceability of the Results} as we are aware that AI techniques are hard to be accepted in the development and certification processes of critical systems, we explicitly consider traceability as a prime citizen in our approach. The idea is to generate automatically also some traceability information to allow the System Analyst to not lose control of the model. To this aim, each time a requirement is matched with a MetaReq, the tool also generates a Requirement Diagram (RD). Fig.~\ref{fig:traceability} reports the RD of the requirement considered in our example. Model elements are reported in the diagram as well, connected to the requirement by means of $<<$satisfy$>>$ stereotyped dependencies; further information (for simplicity in a comment) could be used to specify the role(s) that the model element plays with respect to the requirement.

\paragraph*{Status of the Research}
Currently, the work is focused on clarifying the method and to define a sufficient number of MetaReqs and related MetaFragments. First, we will devise methods to manipulate requirements considering rigid BNF grammars then we will focus on the extraction of lexical information from texts. In particular,  NLTK\footnote{\url{https://www.nltk.org}} and \textsf{spaCy}\footnote{\url{https://spacy.io}} libraries will be used and compared in order to develop the core algorithm of the ClauseExtractor. 
The definition of a SysML Profile that supports the traceability is also on the top of the working agenda.

\section*{Acknowledgment}
\small{The work of Maria Stella de Biase is granted by PON Ricerca e Innovazione $2014 / 2020$ MUR --- Ministero dell'Universit\`a e della Ricerca (Italy) --- with the PhD program XXXVI cycle.}


\begin{thebibliography}{00}
\bibitem{b1} Wynne M. and Hellesoy A., ``The Cucumber Book: Behaviour-Driven Development for Testers and Developers'', Pragmatic Bookshelf, 2012.

\bibitem{b2} North D., ``Introducing BDD,'' in Better Software Magazine, 2006.

\bibitem{b3} Kapurch, S.J., ``NASA Systems Engineering Handbook'', Eds. DIANE Publishing Company, 2010.

\bibitem{b6} Campanile L., de Biase M.S., Marrone S., Raimondo M., Verde L., ``On the Evaluation of BDD Requirements with Text-based Metrics: the ETCS-L3 Case Study.'', 2022, to apper in Proc. of 14th Int.l KES Conference -  Intelligent Decision Technologies, Springer.

\bibitem{b7} Fantechi A., Gnesi S., Ristori G., Carenini M., Vanocchi M., Moreschini P., ``Assisting requirement formalization by means of natural language translation'' in Formal Methods in System Design, 1994.

\bibitem{b8} Buzhinsky I., ``Formalization of natural language requirements into temporal logics: a survey'', In IEEE 17th International Conference on Industrial Informatics (INDIN), vol. 1, 2019.

\bibitem{b9} Sascha K., Cheng B.H.C., ``Automated Analysis of Natural Language Properties for UML Models'', In International Conference on Model Driven Engineering Languages and Systems, 2006.

\bibitem{b10} Brunello A., Montanari A., Reynolds M., ``Synthesis of LTL formulas from natural language texts: State of the art and research directions'', Leibniz International Proceedings in Informatics, Vol. 147, 2019.

\bibitem{b11} Frederiksen S.J., Aromando J., Hsiao M.S., ``Automated Assertion Generation from Natural Language Specifications'' In 2020 IEEE International Test Conference (ITC), 2020.

\bibitem{b12} Brambilla M., Cabot J., Wimmer M., ``Model-driven software engineering in practice.'' Synthesis lectures on software engineering, 2017.

\bibitem{b13} Kharchenko V., Kovalenko A., Siora O., Andrashov A., ``V-models of safety critical system life cycle: Classification and application,'' IEEE 9th International Conference on Dependable Systems, Services and Technologies (DESSERT), 2018.

\bibitem{b14} Burgue{\~n}o L., Claris{\'o} R., G{\'e}rard S., Li S., Cabot J., ``An NLP-based architecture for the autocompletion of partial domain models.'', International Conference on Advanced Information Systems Engineering, 2021.

\bibitem{b15} Furness N., van Houten H., Arenas L., Maarten B., ``ERTMS Level 3: the Game-Changer'', 2017.

\bibitem{b16} Saini R., Mussbacher G., Guo J. L., Kienzle J., ``Automated, interactive, and traceable domain modelling empowered by artificial intelligence'', Software and Systems Modeling, 2022.

\bibitem{b17} Ibrahim M., Ahmad R., ``Class diagram extraction from textual requirements using natural language processing (NLP) techniques'', International Conference on Computer Research and Development, 2010.

\bibitem{b18} Jyothilakshmi M.S., Samuel P., ``Domain ontology based class diagram generation from functional requirements'', International Conference on Intelligent Systems Design and Applications (ISDA), 2012.

\bibitem{b19} L. Burgueño, J. Cabot, M. Wimmer, S. Zschaler; Guest editorial to the theme section on AI-enhanced model-driven engineering (2022) Software and Systems Modeling, 21 (3), pp. 963 - 965.
DOI: 10.1007/s10270-022-00988-0

\end{thebibliography}
\end{document}